\documentclass[elsart,12pt,a4,fleqn,oneside]{article} 
\usepackage{amsmath,amsfonts,graphicx,amsbsy}
\usepackage{euscript}
\usepackage[amssymb]{SIunits}
\usepackage{subfig}
\usepackage{graphicx}
\usepackage{breqn}
\usepackage{bm}
\usepackage{soul}
\usepackage[left]{lineno}
\usepackage{xcolor}
\usepackage{yfonts}
\usepackage{ulem}
\usepackage{yfonts}
\usepackage{tikz-feynman}
\tikzfeynmanset{compat=1.0.0}

\begin{document}
\title{ \bf Possible explanation of not observing  ultra-high energy  cosmic neutrinos}
\author{    J. Rembieli\'{n}ski, J. Ciborowski}
\maketitle

\begin{abstract}
Assuming that neutrinos are spacelike (tachyonic) fermions, we calculate  width  for the kinematically allowed, lepton number conserving,  three-body decay $\nu_{\alpha}\rightarrow \nu_{\alpha} \; \nu_{\beta} \bar{\nu}_{\beta}$   in   the Standard Model.
Decays of tachyonic  neutrinos over cosmological distances can lead to a  reduction  of the neutrino   flux in the   high-energy end of the spectrum. We estimate  upper limits on the spacelike neutrino mass   based on   the PeV-energy cosmological neutrino events  observed  in the IceCube experiment.  These limits  are close to those  deduced from  the measurements  of $m_{\nu}^2$ in the tritium-decay experiment KATRIN.
\end{abstract}


\section{Introduction}


In the past decade a significant effort has been spent on the study  of high energy  astrophysical neutrinos, customarily classified as high-energy (HE) in the range from TeV to 100~PeV and  ultra-high energy  (UHE)  above 100~PeV.
It is widely  accepted  that such neutrinos, produced in distant sources, can travel to Earth undisturbed by the magnetic fields or matter on the way across the Universe, carrying information about the  conditions ruling in the cosmic accelerators.
The IceCube Collaboration~\cite{cite:Aartsen2017a}  discovered high energy extragalactic  neutrinos  in 2013~\cite{cite:Aartsen2013a} whose  energy spectrum,
$ E^2 dN/dE$, was  found similar to that of the $\gamma$-rays~\cite{cite:Ackermann2015a}. The  measured neutrino flux was  initially observed  isotropic (diffuse)   until the discovery of  point-like sources: the blazars  TXS~0506+156~\cite{cite:Aartsen2018a}, PKS B1424-418~\cite{cite:Kadler2016a}, PKS~1502+106~\cite{cite:Lipunov2019a}  and the nearby active galaxy NGC~1068~(Messier~77)~\cite{cite:Aartsen2020a,cite:Abbasi2022b}.
PeV-energy neutrinos have been observed over time~\cite{cite:Aartsen2013b}, including  two most energetic --
a Glashow resonance  candidate  ($\bar{\nu}_e e^-\rightarrow W^-$)~\cite{cite:Aartsen2021a} as well as  just announced  $13\pm5$~PeV muon neutrino event~\cite{cite:Abbasi2023b}.


On the other hand cosmic rays, generated  in cosmic accelerators, have been observed on Earth  with energies  up to about 50~EeV~\cite{cite:Aab2020a,cite:Abraham2008a,cite:Anchordoqui2019}.  This limit  corresponds to the  cut-off value for protons interacting with photons of the cosmic microwave background radiation~\cite{cite:Greisen1966,cite:Zatsepin1966} but  can also be explained  as a consequence of a maximal energy reachable  in cosmic  sources. Since UHE neutrinos  are expected to carry  $3 \div 5$~\% of the primary  hadron energy, they  should be observed on Earth
with  energies even up to a few~EeV~\cite{cite:Berezinsky1969,cite:Strecker1979,cite:Hill1983,cite:Yoshida1993}.
However this is not the case so   the question  of  not observing   UHE neutrinos is still open.
Lorentz symmetry violation has been  considered in this context~\cite{cite:Coleman1997,cite:Gorham2012,cite:Anchordoqui2014},
involving   a kinematical  high-energy cutoff, possibly leading to  non-observation of neutrinos beyond a certain energy.
According to  another concept,  UHE primary and secondary charged particles,  spiraling    in a   magnetic field  in the source, are subject to a significant radiative energy loss prior to eventually  decaying into neutrinos.
Also  sensitivity of the present detectors to neutrinos in the UHE range can be  insufficient.

In the present paper we adopt a  hypothesis that  neutrinos are spacelike (thus superluminal)  fermions
and  consider kinematically allowed   neutrino decays to explain the reduction of the  neutrino  flux in the   high energy end of the spectrum.
A tachyonic particle is characterised by a spacelike energy-momentum dispersion relation, $E^2 - \bm{k}^2 = -\kappa ^2$, where $\kappa$ denotes the tachyonic mass, in contrast to the inertial  mass, $m$,  satisfying  the relation $E^2 - \bm{p}^2 = m ^2$ for massive particles.
We derive the neutrino decay width  within the framework of  a slightly  modified  Standard Model in the neutrino sector and translate it into  the decay probability in the expanding Universe. We determine  upper limits on the tachyonic neutrino mass, $\kappa$,  from the observation on Earth of the  highest energy    cosmological  neutrino events   and  find  these results very close to  the independent measurement of $m_{\nu}^2$ in   the tritium decay experiment KATRIN.
In the following   the  term  "neutrino"  implicitly  refers to  the "spacelike neutrino"  unless stated otherwise.

\section{Why spacelike neutrinos}

Neutrinos have been associated  with   superluminality since decades. The initial underlying justification  consisted in repeated occurrences of negative  or consistent with negative central values for the electron antineutrino mass squared, observed  in numerous tritium decay experiments, to quote only the most recent:
$m^2= -0.6 \pm 2.2{(\rm stat.)} \pm 2.1{\rm (syst.)}$ ~\cite{cite:Kraus2005}~(Mainz Collaboration, 2005) and
$m^2= -0.67 \pm 2.53{(\rm tot.)}$ ~\cite{cite:Aseev2011}~(Troitsk Collaboration, 2011). These results were  superseded   by those obtained in the  presently running  experiment KATRIN which  achieved an unprecedented accuracy. Their first measurement again  yielded a negative central value of the mass squared,  $m_{\nu}^2=-1.0^{+0.9}_{-1.1}$~eV$^2$~\cite{cite:Aker2019}~(2019), while the following  measurement period ended with  a positive central value, $m_{\nu}^2=0.26\pm 0.34$~eV$^2$~\cite{cite:Aker2022}~(2022), however  consistent with being negative within even less than $1\sigma$. In contrast to any other  elementary object,  letting aside the common  prejudice,  one must accept that according to  the present evidence, the  four-momentum of the neutrino may as well be  spacelike.
Admitting the above  requires  providing  an adequate  theoretical  description of spacelike neutrinos at the quantum field theory level.

Early attempts to describe tachyonic neutrinos within  the standard (Einsteinian) relativity were unsuccessful at any level and did not lead to a solution of  the essential problems within this framework, like causality violation, negative energies or vacuum instability. It was shown already half a century ago that  the standard  relativistic quantum field theory is inapplicable for describing spacelike particles~\cite{cite:Kamoi1971},\cite{cite:Nakanishi1972}.
The first proposition of a Dirac-like equation for spin-$\frac{1}{2}$  spacelike  neutrinos is due to Chodos et al.~\cite{cite:Chodos1985}, although his formalism was not unitary. However  a turning point in these studies came about when it was realised  that  spacelike particles can be causally described using a special procedure of clock synchronisation~\cite{cite:Rembielinski1997}. This  modification affects uniquely the superluminal sector, leaving the subluminal sector unchanged  since different  clock synchronisation schemes lead to equivalent results in the latter  case. This particular procedure implies existence of  a preferred frame of reference  which plays a role  only for the superluminal sector.  It  does   not overrule  the  validity of the relativity principle and the Lorentz symmetry  in  the subluminal sector as well as respects the Lorentz covariance of the entire theory.   This seemingly minor modification  allows to construct a  Lorentz-covariant quantum field-theoretical model of a relativistic helicity $-\frac{1}{2}$ tachyonic fermion~\cite{cite:Rembielinski1997} and avoid  known  fundamental difficulties related to spacelike particles. In a recent  paper we formulated a consistent quantum field theory of the spacelike neutrino with both  $\pm \frac{1}{2}$ helicity components, within the   formalism based on the existence of a preferred frame, affecting the neutrino sector only~\cite{cite:Rembielinski2021a}.
A further motivation to consider neutrinos as spacelike particles is of a theoretical nature.  One thread  stems from the fact that, according to our new results the two-helicity  neutrino state for $E \approx \kappa $ reduces to one-helicity state  for $E \gg \kappa$, as observed in nature; another one is  that the  property of the neutrino known for decades --  separate $C$ and $P$ violation --  would follow from the fact of their spacelike nature~\cite{cite:Rembielinski2021a}.
The notion of the  preferred frame  has scored   numerous   references  in the context of the quantum theory.
A natural candidate to consider is  the    Cosmic Neutrino Background (CNB) frame,  an artefact of the electroweak phase
transition~\cite{cite:Baumann2019}, defined as  a  local  reference frame in which the CNB is isotropic.
According to cosmological predictions, the CNB frame should practically coincide with cosmic microwave background (CMB) radiation frame  in which the microwave background radiation is isotropic.

Although all relevant details can be found in the cited papers~\cite{cite:Rembielinski1997,cite:Rembielinski2021a}, we point here to two most
serious difficulties of the standard tachyonic theory -- vacuum instability and the causality problem, which
are frequently passed over in numerous  publications. These problems disappear in our approach
due to  the  assumption of existence of the preferred frame, identified with the  CNB frame.  Let us denote the four-velocity of the CNB frame as seen by an observer
by  $u^{\mu}$, where  $u^2=1$, and the  four-momentum of a free particle by $k^{\mu}$.  Then $q=u^{\mu} k_{\mu}$ is a Lorentz invariant so
\begin{linenomath}\begin{equation}\label{eq:SpectralCondition9}
  q>0
\end{equation}\end{linenomath}
is an  invariant spectral condition,  irrespective of the  dispersion relation of the considered particle  (timelike, lightlike or spacelike). In all three cases, for observers  in the preferred frame, $u^0=1,\bm{u}=0$ and $k^{\mu}=(E,\bm{k})$, by means of this condition,  the energy of a particle is positive.
In the two former cases, it is positive for observers in all inertial frames because the upper (physical) parts of the two-sheet energy-momentum  hyperboloid  or the
energy–momentum cone transform into themselves under the action of the Lorentz group. In the spacelike case, the one-sheet four-momentum hyperboloid  is as well
divided by  the Lorentz invariant condition $q>0$  into the upper (physical) and the lower (unphysical) parts. This condition fixes  a lower bound of energy in any
inertial frame, necessary to avoid  vacuum instability and causality violation  as well as  to carry out a proper quantisation procedure of spacelike
fields~\cite{cite:Rembielinski2021a}. Vacuum instability stems from the fact of the appearance of negative energies without a lower bound and a possibility of a
spontaneous creation from the vacuum of tachyon-antitachyon pairs with opposite four-momenta, $k^{\mu}$ and $-k^{\mu}$,   explicitly satisfying  energy-momentum
conservation. In the approach with the preferred frame, the respective invariants  are $uk$ and $-uk$,    so both the tachyon and antitachyon in a given pair  cannot
simultaneously satisfy~(\ref{eq:SpectralCondition9}). Therefore, the  condition $q>0$ rules out  the possibility of vacuum instability. We note that this condition is
analogous to choosing the upper energy-momentum hyperboloid (or cone) as the physical one for massive (or massless) particles, respectively.
Now, as regards  causality,  identification of the preferred frame  with the CNB  frame allows to interpret  the cosmic time, measured by a physical clock moving along
the Hubble flow, as   the absolute time determining   causality relations between events.

For more  contributions to the tachyonic neutrino hypothesis one is referred to abundant
literature~\cite{cite:Chodos1992,cite:Caban2006,cite:Ehrlich2015,
cite:Somogyi2019,cite:Schwartz2016}. In the following  we use  $c=1$ but we preserve $\hbar$ for clarity.

\section{Decays  of spacelike neutrinos}

\subsection{Decay width}

A unique  property of the  tachyonic neutrino  is  its  decay into the neutrino of the same flavour   and an additional  state,  $\nu_{\alpha} \rightarrow \nu_{\alpha} + X$ -- a process  which is  kinematically allowed  under the  conservation of  four-momentum and respecting  the spacelike energy-momentum dispersion relations while    it is forbidden  for  massive neutrinos.
The two dominant channels  are  the three-body decay  $\nu_{\alpha}\rightarrow \nu_{\alpha} \; \nu_{\beta} \bar{\nu}_{\beta}$   and  the  radiative decay $\nu_{\alpha}\rightarrow \nu_{\alpha} \gamma$, where the indices $\alpha,\beta$  run over three flavours,   $e,\mu,\tau$. These processes have already been preliminarily  analysed, albeit  in a different context~\cite{cite:Caban2006}.
Also the  following decay channel, $\nu \rightarrow \nu \,e^+e^-$, has  been  considered in literature, however we do not deal with this process as it is   forbidden in our approach due to violating the spectral condition~(\ref{eq:SpectralCondition9}) (see below).

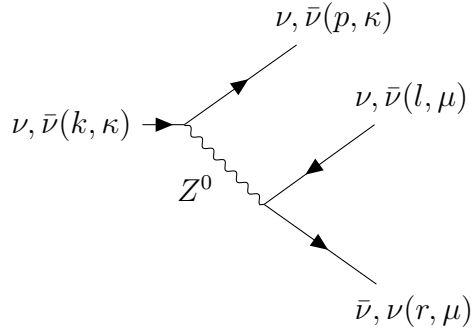
\begin{figure}[!htbp]
    \begin{center}
        \begin{tikzpicture}
\begin{feynman}
\vertex (a) {\(\nu,\bar{\nu}(k,\kappa)\)};
\vertex [right=of a] (b);
\vertex [above right=of b] (f1) {\(\nu,\bar{\nu}(p,\kappa)\)};
\vertex [below right=of b] (c);
\vertex [above right=of c] (f2) {\( \nu,\bar{\nu}(l,\mu)\)};
\vertex [below right=of c] (f3) {\(\bar{\nu},\nu(r,\mu)\)};
\diagram* {
(a) -- [fermion] (b) -- [fermion] (f1),
(b) -- [boson, edge label'=\(Z^{0}\)] (c),
(c) -- [anti fermion] (f2),
(c) -- [fermion] (f3),
};
\end{feynman}
\end{tikzpicture}
    \caption{\label{fig:Graph1} Feynmann diagram of the tachyonic neutrino/antineutrino  three-body decay. The neutrino momenta and tachyonic masses that appear in Eq.~\ref{eq:dGamma2} are indicated in brackets.}
    \end{center}
\end{figure}

Below  we present calculations of  the amplitudes and the widths  for   the three-body decay, depicted in Fig.~\ref{fig:Graph1},  within the full framework of  the aforementioned   quantum field theory  of spacelike neutrinos~\cite{cite:Rembielinski2021a}.  We do not deal with the radiative process since its  width is  many orders of magnitude smaller, compared to that for the three-body channel.

The width for the process shown in Fig.~\ref{fig:Graph1} can be calculated  from the formula~\cite{cite:Rembielinski2021a}
\begin{linenomath}\begin{equation}\label{eq:dGamma1}
 d\Gamma = \frac{G_F^2 |M|^2 d\Phi}{(2\pi)^5 k^0 \sqrt{|\bm{k}|}},
\end{equation}\end{linenomath}
where  $d\Phi = \theta (up) \theta (ul) \theta (ur)\delta^4 (k-p-l-r) \delta (p^2+\kappa^2) \delta (l^2+\mu^2) \delta(r^2+\mu^2)d^4 p\, d^4 l\, d^4 r$ is the phase-space element. The matrix element squared, $|M|^2$, has the following general form
\begin{linenomath}\begin{multline}\label{eq:dGamma2}
|M|^2  = \frac{64}{ \sqrt{(uk)^2+\kappa^2}\sqrt{(up)^2+\kappa^2} \sqrt{(ul)^2+\mu^2} \sqrt{(ur)^2+\mu^2}    }\\
\Bigg  (
\kappa^4\mu^4 + \kappa^4\Big[ \mu^2 \Big((ul)^2+(ur)^2\Big) +(ul)^2(ur)^2\Big] + \mu^4\Big[ \kappa^2 \Big((up)^2+(uk)^2\Big) +(up)^2(uk)^2\Big] \\
+ \kappa^2\mu^2\Big[ (kr) \Big((uk)(ur)-(up)(ul)\Big) +(up)^2(ur)^2 + (uk)^2(ul)^2\Big]\\
+ (kr) \Big((uk)(ul)-(up)(ur)\Big)  \Big(\kappa^2 (ul)(ur)- \mu^2(uk)(up)\Big)
- (uk) (up)  (ur) (ul) (kr)^2
\Bigg ),
\end{multline}\end{linenomath}
where $u$, as above,  denotes the four-velocity of the preferred  frame (CNB)  and the scalar products are of the form  $uk=u^{\lambda}k_{\lambda}$ etc.
Since below we  calculate the width for the decay in the  preferred  frame (CNB),  we  put  $u=(1,0,0,0)$, for which $uk=k^0=E$.
In view of the complexity of these calculations, the width $\Gamma$ cannot  be easily derived  as a closed form expression. Instead, we  have developed a simple effective formula on the basis of dimensional and numerical analysis, accurate to the 8-th decimal in  the energy range under study (TeV-PeV)
\begin{linenomath}\begin{equation}\label{eq:Effective1}
 \Gamma (E,\kappa,\mu) = f\; n_f\,  G_F^2 (\kappa^4 + 4 \mu^4)E,
\end{equation}\end{linenomath}
where $f=\frac{5}{18 (2\pi)^3}$ and $n_f=3$  accounts for the three neutrino flavours in  the final state of the  $\nu\bar{\nu}$ pair (cf. Fig.~\ref{fig:Graph1}).
Understandably,  the powers of the masses and energy, integers to a very high precision,  add up to five. We note the strong dependence of the width on the neutrino masses as well as the dominating weight of the  mass  $\mu$. The numerical calculations require a very high working precision  to obtain a numerically stable result.

Given the mass differences squared measured in oscillation experiments, one can derive the masses of the heavier mass states as a function of the mass of the lowest mass state, $m_1$:  $  m_2(m_1)=\sqrt{\Delta m_{21}^2  + m_1^2}$ and $  m_3(m_1)=\sqrt{\Delta m_{21}^2 + \Delta m_{32}^2 + m_1^2}$,
where $\Delta m_{21}^2 = 7.4\times 10^{-5}$~eV$^2$  and  $\Delta m_{32}^2 = 2.5\times 10^{-3}$~eV$^2$~\cite{cite:NuFit} (using the conventional labeling of the neutrino masses). It has been shown   that the oscillation pattern of  tachyonic neutrinos is the same as that of the massive ones~\cite{cite:Caban2006a}, i.e., the above formulae  relating the  mass states are  applicable for tachyonic neutrinos too when the conventional masses, $m_i$, are replaced by the tachyonic masses,  $\kappa_i$. If the mass of the lowest neutrino state is of the order of a fraction of an eV,  the above differences of mass squares  can be neglected and the same given mass, $\kappa$, can be assigned to all three flavours, implying also $\kappa \approx  \mu$.
This is true in particular  for  the value derived from the latest  measurement of KATRIN, $m_{\nu}^2=0.26\pm 0.34$~eV$^2$~\cite{cite:Aker2022},  from which one can determine the lower limit, $m_{\nu}^2 > -0.3$~eV$^2$,  at a 90\% c.l. and the corresponding  upper limit on  the tachyonic neutrino mass, $\kappa < 0.55$~eV, below referred to as the KATRIN limit.
In the  special relativity  framework the mean lifetime for the decay of a given tachyonic neutrino  can obtained from the well known relation $\tau = \hbar /\Gamma$, valid in the CNB  frame of reference,  where  the  width for the decay has already been accounted for three possible final states.
Simply put,  the mean lifetime calculated this way  would describe   the neutrino moving  through a static Universe, with the conventional understanding of time and  distance.  Obviously, in view of  cosmologic distances between neutrino sources and the detector on Earth, it is indispensable to insert  the decay width~(\ref{eq:Effective1})  into  the environment of the expanding Universe, in particular consider the decay according to  the flow of the cosmic time,  which requires a description involving  the  neutrino energy  at the source  of emission as well as the   redshift of the source.

Lastly, the following remarks regarding the process $\nu \rightarrow \nu \,e^+e^-$ are in order. Cohen and Glashow admitted that this decay  was forbidden within the standard Einsteinian relativity~\cite{cite:Cohen2011}.
In order to demonstrate that it is also forbidden within the  preferred  frame  framework,  it is sufficient to consider  this decay in the center-of-mass frame of the $e^+ e^-$ pair and apply the spectral condition~(\ref{eq:SpectralCondition9}).

\subsection{Survival probability in the expanding Universe}\label{subsec:Survival}

Since we consider decays of  spacelike neutrinos, we elaborate on this nonstandard case in cosmology in detail (we note in passing that the first attempt of describing tachyon  kinematics  in cosmology is  owed to Narlikar and Sudarshan~\cite{cite:Narlikar1976}).
According to the commonly used  $\Lambda$CDM  model~\cite{cite:Workman2022}, the geometry of our Universe is described by the Friedmann-Lema\^{i}tre-Robertson-Walker (FLRW) space-time~\cite{cite:Dodelson2020}.
Since our Universe  is flat at a large scale,  we put the curvature parameter $k=0$ and  the corresponding FLRW  line element takes the form
\begin{linenomath}\begin{equation}\label{eq:LineElement2}
ds^2 =  dt^2 - a(t)^2 \Big (  dr^2  + r^2 \left( d\theta^2 + \sin^2\theta d\varphi^2 \right )  \Big ).
\end{equation}\end{linenomath}
The scale factor $a(t)$ is related to the redshift, $z$, through the formula $a(t)=(1+z)^{-1}$ .
The motion of the spacelike neutrino in the flat FLRW spacetime is governed by the spacelike geodesics and the corresponding dispersion relation. In the spacelike case,  $ds^2<0$,  we can parametrise the line element  in terms of an affine parameter, $\lambda$,   $ds^2=- d\lambda^2$  and in the consequence the FLRW metric takes the form
\begin{linenomath}\begin{equation}\label{eq:gmunu1}
g_{\mu\nu} dx^{\mu}dx^{\nu} = - d\lambda^2.
\end{equation}\end{linenomath}
Defining the four-momentum, $k^{\mu}$, in the standard way as
\begin{linenomath}\begin{equation}\label{eq:MomentumDef1}
k^{\mu}=\kappa \,dx^{\mu}/d\lambda,
\end{equation}\end{linenomath}
where  $\kappa$ is the particle mass, we obtain the dispersion relation  in the form
\begin{linenomath}\begin{equation}\label{eq:gmunu2}
g_{\mu\nu} k^{\mu}k^{\nu} = -\kappa^2.
\end{equation}\end{linenomath}
In the case of a flat FLRW one has
\begin{linenomath}\begin{equation}\label{eq:LineElement3}
(k^0)^2 - a(t)^2 \Big (  (k^r)^2  + r^2 \left( (k^\theta)^2 +  (k^\varphi)^2 \sin^2\theta \right )  \Big ) = -\kappa^2
\end{equation}\end{linenomath}
so taking into account that
\begin{linenomath}\begin{equation}\label{eq:LineElement4}
g_{ij} k^i k^j= a(t)^2 \Big (  (k^r)^2  + r^2 \left( (k^\theta)^2 + (k^\varphi)^2 \sin^2\theta \right )  \Big ) \equiv \bm{k}^2
\end{equation}\end{linenomath}
is the  momentum squared, we can rewrite~(\ref{eq:LineElement3}) in the standard form
\begin{linenomath}\begin{equation}\label{eq:DispersionRelation1}
(k^0)^2 - \bm{k}^2 = -\kappa^2.
\end{equation}\end{linenomath}
Now, we  can simplify the  considerations by taking into account the fact  that the direction of the spacelike neutrino is unchanged throughout its motion.
This implies $d\theta d\varphi=0$ and the particle line element~(\ref{eq:gmunu1}) reduces to the form
\begin{linenomath}\begin{equation}\label{eq:DispersionRelation2}
-d\lambda^2= dt^2-a(t)^2 dr^2.
\end{equation}\end{linenomath}
In such a case $k^{\theta}=k^{\phi}=0$ and $\bm{k}^2=a(t)^2 (k^r)^2$.
Substituting~(\ref{eq:MomentumDef1}) into the geodesic equations
\begin{linenomath}\begin{equation}\label{eq:Geodesic0}
 \frac{d^2 x^{\mu}} {d\lambda^2} + \Gamma^{\mu}_{\alpha\beta} \frac{d x^{\alpha}}{d\lambda} \frac{d x^{\beta}}{d\lambda}=0
\end{equation}\end{linenomath}
yields straightforwardly
\begin{linenomath}\begin{equation}\label{eq:Geodesic1}
\kappa \frac{d k^{\mu}} {d\lambda} + \Gamma^{\mu}_{\alpha\beta} k^{\alpha}k^{\beta}=0
\end{equation}\end{linenomath}
and by calculating the connection coefficients for this case we can reduce~(\ref{eq:Geodesic1}) to  only one independent equation of the form
\begin{linenomath}\begin{equation}\label{eq:Geodesic2}
k^0 dk^0 +  |\bm{k}| \frac{da}{a}=0.
\end{equation}\end{linenomath}
Therefore the geodesic motion of the spacelike neutrino in a flat FLRW spacetime is determined by the dispersion relation~(\ref{eq:DispersionRelation1}) and the geodesic condition~(\ref{eq:Geodesic2}). Solving the system of these two equations  one  obtains
\begin{linenomath}\begin{eqnarray}\label{eq:Solution1}
(k^0)^2 &=&   C^2 a^{-2} -\kappa^2\\
|\bm{k}| &=& Ca^{-1}.
\end{eqnarray}\end{linenomath}
The constant $C$ can be determined by  assuming that the  emission of the particle with energy $E_e$  took place  in the epoch characterised  by the redshift $z_e$
\begin{linenomath}\begin{equation}\label{eq:ConstantC}
C^2 = \frac{E_e^2+\kappa^2  }{ (1+z_e)^2},
\end{equation}\end{linenomath}
using the  aforementioned relationship between the scale factor  $a$ and $z$.
Finally one obtains the following expression for the neutrino energy, $E$, in the epoch determined by the redshift $z$
\begin{linenomath}\begin{equation}\label{eq:Solution3}
E  = \sqrt{ (E_e^2+\kappa^2 ) \left ( \frac{1+z}{1+z_e}\right )^2 -\kappa^2}.
\end{equation}\end{linenomath}
In particular, $z=0$ corresponds to an observer on Earth in the present epoch.
Eq.~\ref{eq:Solution3} states that the energy $E$  is smaller than  that with which the neutrino was emitted, $E_e$. Neglecting the tachyonic neutrino  mass terms, $\kappa ^2$, compared to  the energy of emission,  $E_e ^2$,  one obtains the following relation
\begin{linenomath}\begin{equation}\label{eq:Solution4}
E =  E_e \frac{1+z}{1+z_e}.
\end{equation}\end{linenomath}
The velocity of a neutrino  emitted with TeV or higher  energies only infinitesimally exceeds  the velocity of light in vacuum,  $c$.
Spacelike neutrinos have positive energies ($E>0$) in any local  reference frame  in which the CNB  is isotropic~\cite{cite:Rembielinski2021a}.

A  neutrino moving through a  homogeneous,  expanding Universe "experiences"  the cosmic  time.
Our  goal is to determine the neutrino survival probability in terms of the cosmic time elapsing  during  its travel from the point of  emission  to the point of detection, parametrised by $z_e$  and $z$, respectively  ($z=0$  on Earth in the present epoch). The standard procedure lies in the relationship of the survival  probability differential,  $dp_{\rm s}(z)$, with the cosmic time differential,  $dt(z)$, identified with the cosmic distance differential travelled with the velocity of light,  which has the form
\begin{linenomath}\begin{equation}\label{eq:CosmicProb1}
\frac{dp_{\rm s}(z)}{p_{\rm s}(z)} = - \frac{\Gamma(E)}{\hbar} dt(z),
\end{equation}\end{linenomath}
where  the decay  width $\Gamma(E)$ is given by~(\ref{eq:Effective1})  but with the energy $E$  replaced by the r.h.s. of  Eq.~\ref{eq:Solution3}.
In the $\Lambda$CDM model  the cosmic time differential is given by
\begin{linenomath}\begin{equation}\label{eq:CosmicTime1}
dt(z) = - \frac{dz}{(1+z) H(z)},
\end{equation}\end{linenomath}
with the function $H(z) $  defined  as
\begin{linenomath}\begin{equation}\label{eq:HubbleParameter1}
H(z) = H_0 \sqrt{\Omega_r (1+z)^4+ \Omega_m (1+z)^3 + \Omega_k (1+z)^2 + \Omega_{\Lambda}},
\end{equation}\end{linenomath}
where  $H_0$  is the Hubble constant, and the  normalised energy densities for photons and  neutrinos, baryons and  dark matter, and  dark energy  are denoted $\Omega_r$,  $\Omega_m$ and   $\Omega_{\Lambda}$, respectively,     while  $\Omega_k = 1 - (\Omega_r + \Omega_m + \Omega_{\Lambda})$  determines a deviation from flatness.
According to observations, $\Omega_r + \Omega_m + \Omega_{\Lambda}=1$ so   $\Omega_k=0$. Moreover,  $\Omega_r$ is very small in our epoch so it can be neglected and consequently  $\Omega_m = 1 - \Omega_{\Lambda}$.
Therefore the cosmic time differential~(\ref{eq:CosmicTime1}) takes the form
\begin{linenomath}\begin{equation}\label{eq:CosmicTime2}
dt(z) = - \frac{dz} { (1+z)  H_0 \sqrt{     (1 - \Omega_{\Lambda} )  (1+z)^3 + \Omega_{\Lambda}}   },
\end{equation}\end{linenomath}
where $H_0=67.5$~km s$^{-1}$ Mpc$^{-1}$  and $\Omega_{\Lambda}=0.685$.
Integrating~(\ref{eq:CosmicTime2}) over $z$ yields the cosmic time interval between  the emission time at $z_e$  and the detection time at $z$
\begin{linenomath}\begin{multline}\label{eq:CosmicTime3}
t(z_e,z) =  \int_{z_e}^{z} dt(z^{\prime}) = \\
 \frac{1}{3 H_0 \sqrt{\Omega_{\Lambda}}}
\ln  \Bigg [
\Bigg ( \frac{\sqrt{\Omega_{\Lambda}} +\sqrt{(1+z)^3 (1-\Omega_{\Lambda})  +\Omega_{\Lambda} }   } {\sqrt{\Omega_{\Lambda}} -\sqrt{(1+z)^3 (1-\Omega_{\Lambda})  +\Omega_{\Lambda} }   }  \Bigg ) \; \Bigg (
\frac{\sqrt{\Omega_{\Lambda}} -\sqrt{(1+z_e)^3 (1-\Omega_{\Lambda})  +\Omega_{\Lambda} }   } {\sqrt{\Omega_{\Lambda}} +\sqrt{(1+z_e)^3 (1-\Omega_{\Lambda})  +\Omega_{\Lambda} }   } \Bigg )
\Bigg ] .
\end{multline}\end{linenomath}
Note that the emission time $t_e = t(z_e,z_e)=0$  while the detection time on Earth in the present epoch is given by  $t_d=t(z_e,0)$ which, up to the factor $c$, corresponds to the distance between  the source and   Earth.
Now, by means of~(\ref{eq:Effective1}) and~(\ref{eq:Solution3}) one obtains
\begin{linenomath}\begin{equation}\label{eq:CosmicProb2}
\frac{dp_{\rm s}(z)}{p_{\rm s}(z)} =  \frac{ f\, n_f\; G_F^2 (\kappa^4 + 4 \mu^4) \sqrt{-\kappa^2  + (E_e^2+\kappa^2  ) \left ( \frac{1+z}{1+z_e} \right )^2} }  { \hbar  (1+z)  H_0 \sqrt{     (1 - \Omega_{\Lambda} )  (1+z)^3 + \Omega_{\Lambda}}    } \,dz.
\end{equation}\end{linenomath}
Neglecting $\kappa^2$  under  the square root yields  the cancellation of the $(1+z)$ terms in the numerator and the denominator and
both sides  of Eq.~\ref{eq:CosmicProb2} can then be integrated analytically.
As a result we obtain the following expression for the neutrino decay probability
\begin{linenomath}\begin{equation}\label{eq:CosmicProb3}
p_{\rm s} = \exp{\Bigg ( -f\,n_f\; G_F^2 \Big ( \kappa^4 + 4 \mu^4 \Big )\, \frac{E_d\, \tau(z_e,z)}{\hbar} \Bigg ) }
\end{equation}\end{linenomath}
where $E_d$ is the neutrino energy measured presently on Earth, obtained from~(\ref{eq:Solution4}) for  $z =0$.
In the following we use the approximation $\kappa\approx \mu$, justified above. The function $\tau(z_e,z)$
is given by the following integral
\begin{linenomath}\begin{equation}\label{eq:FunctionTau1}
\tau(z_e,z) =  \frac{1}{H_0} \int_{z}^{z_e} \frac{dz^{\prime}}{ \sqrt{     (1 - \Omega_{\Lambda} )  (1+z^{\prime})^3 + \Omega_{\Lambda}} } = W(z_e) - W(z),
\end{equation}\end{linenomath}
expressed in terms of  the hypergeometric function $_2F_1$
\begin{linenomath}\begin{equation}\label{eq:FunctionTau2}
W(z)= \frac{ (1+z)   } {H_0 \sqrt{\Omega_{\Lambda}}   } \;   _2F_1 \left( \frac{1}{3},\frac{1}{2},\frac{4}{3}, \frac{(1+z)^3 (\Omega_{\Lambda}-1) }{\Omega_{\Lambda}} \right ).
\end{equation}\end{linenomath}
Again, for an observer on Earth  we  put $z=0$ in~({\ref{eq:FunctionTau1}}).

\section{Results and discussion}

\begin{figure}[!htbp]
    \begin{center}
        \includegraphics[width=0.9\linewidth]{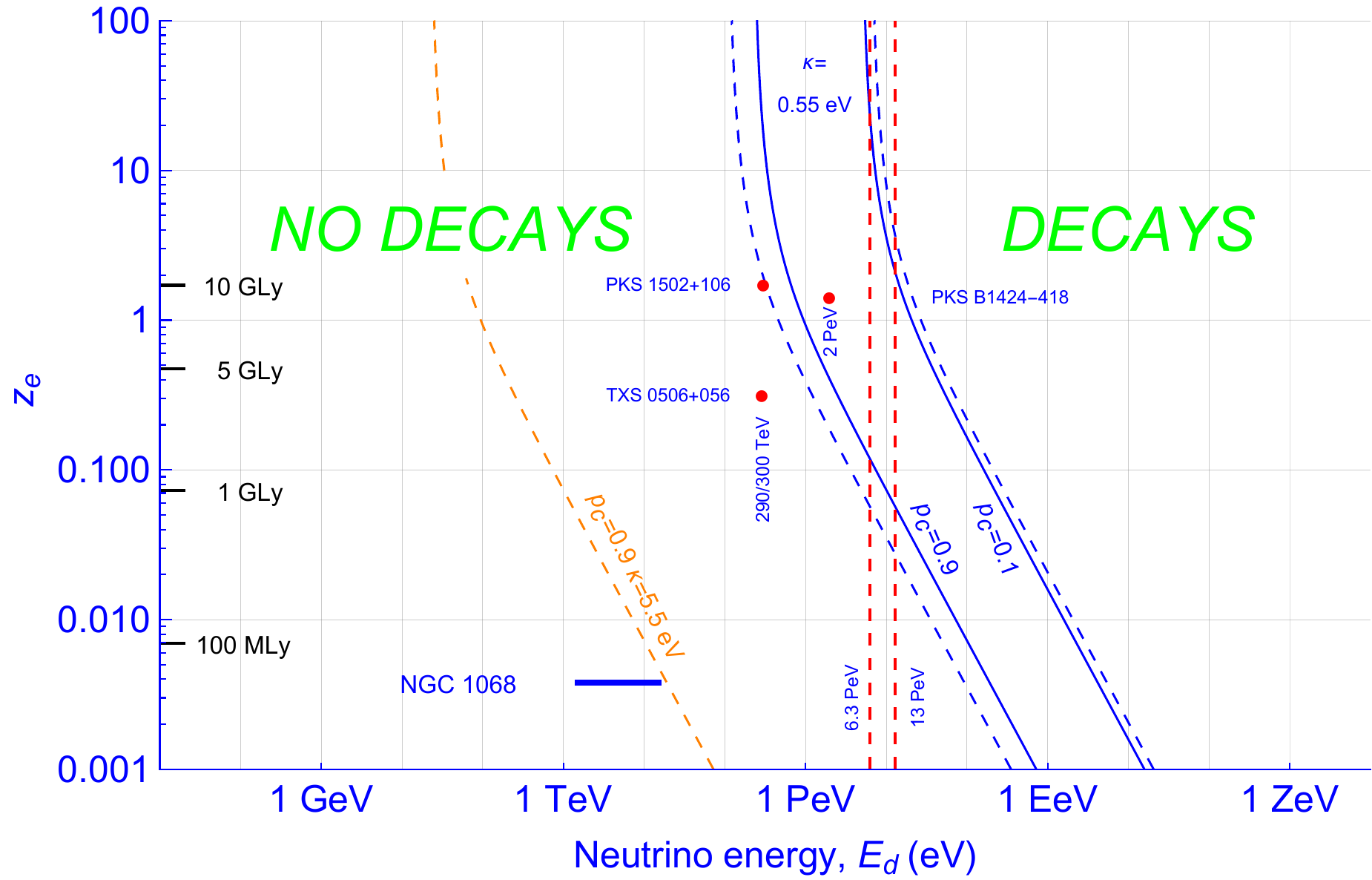}
    \caption{\label{fig:Fig6} Solid curves mark contours corresponding to $p_{\rm c}=0.9$ or $p_{\rm c}=0.1$, as indicated, where $p_{\rm c}$ is the neutrino  survival probability assumed on Earth~(\ref{eq:CosmicProb4}),  for  $\kappa=0.55$~eV (KATRIN limit); the accompanying dashed curves are drawn for  $p_{\rm c}=0.95$ and $p_{\rm c}=0.05$, correspondingly. The separate  dashed curve shows the prediction for an upper limit of approximately $\kappa=5.5$~eV ($p_{\rm c}=0.9$), based on observing  high energy neutrinos from the nearby galaxy  NGC~1068 ($z_e=0.0038$) alone (the horizontal  solid line indicates  the corresponding energy range.
}
    \end{center}
\end{figure}

Given the formula~(\ref{eq:CosmicProb3}), we  analyse the survival  probability   of  tachyonic neutrinos in terms of the variables  $E_e$, $\kappa$ and $z_e$, keeping in mind the relation of the energy with which the neutrino was emitted from the source, $E_e$  and the energy observed on Earth, $E_d$.
We also discuss solutions of  Eq.~\ref{eq:CosmicProb3} on Earth ($z=0$)   for a preset   value of  the probability,  $p_{\rm c}=$const
 \begin{linenomath}\begin{equation}\label{eq:CosmicProb4}
p_{\rm s} = p_{\rm c},
\end{equation}\end{linenomath}
which defines   what   part of the flux emitted from a given  source   has survived the   travel time  to Earth.
We have adopted a working  convention  that the neutrino flux  reaches the Earth  almost unaffected  when  $p_{\rm c}>0.9(0.95)$  and  almost entirely vanishes when  $p_{\rm c}<0.1(0.05)$, therefore  the intermediate  range between the two values  corresponds  to  fluxes non-negligibly  depleted due to  decays
(the number of observed events additionally depends on the cross-section and detector characteristics).
Since in the expanding  Universe  the proper variable standing for the distance of a source is its  redshift, $z_e$, we show in Fig.~\ref{fig:Fig6}    constant probability curves  on a $z_e$  vs.  $E_d$ plot,  for a fixed value of  $\kappa =0.55$~eV (KATRIN limit). 

In order to compare the predictions with measurements,   we consider  five neutrino events observed  by  the IceCube Collaborations:
($i$)  290~TeV  from  the blazar TXS 0506+056 at $z_e$=0.3365 (3.9~Gly);
($ii$)  300~TeV  from  the blazar PKS 1502+106 at $z_e$=1.839 (10.2~Gly);
($iii$) 2~PeV from the blazar PKS B1424-418 at $z_e$=1.522 (9.6~Gly);
($iv$) $6.05\pm0.72$~PeV event identified as a Glashow resonance (referred to as a 6.3~PeV event), from an unidentified source;
($v$)  $13\pm5$~PeV  recently published event,  from an unidentified source.
The three events from identified sources   are marked by full points on the $z_e$ vs. $E_d$ plot (Fig.~\ref{fig:Fig6}).

\begin{figure}[!htbp]
    \begin{center}
        \includegraphics[width=0.9\linewidth]{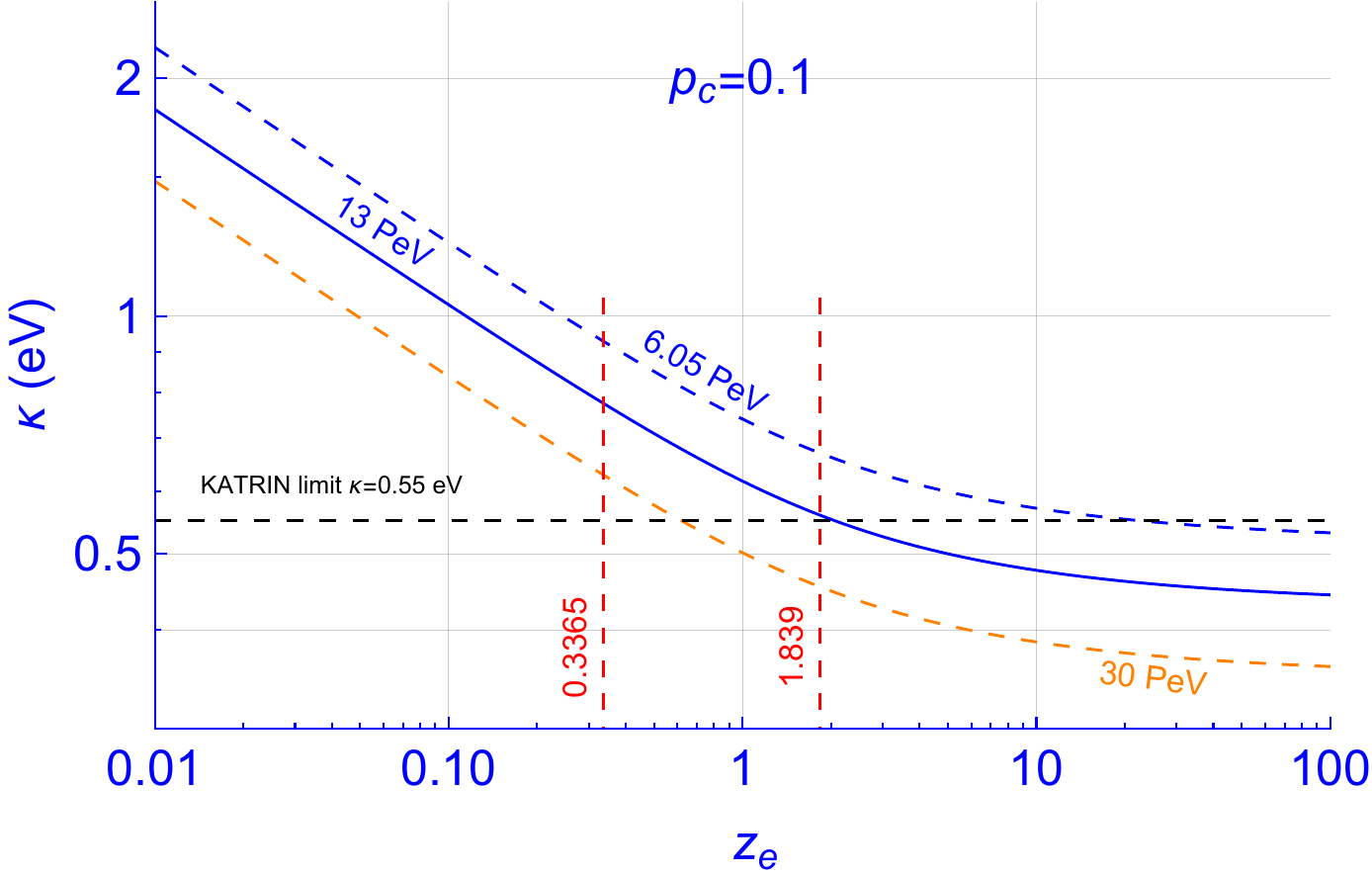}
    \caption{\label{fig:Fig3} Contours of fixed  neutrino  survival probability $p_{\rm c}=0.1$,  for  $E_d=6.3$~PeV (Glashow resonance), 13~PeV (the highest energy event of IceCube) and 30~PeV for comparison.}
    \end{center}
\end{figure}

Observing one neutrino event from a  source with a known $z_e$  allows to roughly  estimate an independent  upper  limit on the neutrino mass  by 
solving~(\ref{eq:CosmicProb4}) for $\kappa$,  under an assumption  of the survival probability, $p_{\rm c}$,   which brings in an element of  uncertainty.  Taking    $p_{\rm c}=0.1$   yields a conservative upper bound  $\kappa < 2$~eV,  $\kappa < 1.4$~eV  and $\kappa<0.9$~eV  for the blazars ($i$)--($iii$), respectively. For  events with  unknown redshifts  we adopt  $z_e= 0.3365$  or $z_e=1.839$ as an example.  And thus for the Glashow resonance event  we obtain $\kappa < 0.9$~eV and $\kappa < 0.7$~eV, respectively,   whereas  for the 13~PeV event  $\kappa < 0.8$~eV and $\kappa < 0.6$~eV.
The corresponding constant-probability contours  on a  $\kappa$  vs. $z_e$  plot are   shown in Fig.~\ref{fig:Fig3} for $p_{\rm c}=0.1$  and   the following energies: $E_d=6.3$~PeV (Glashow resonance), the highest energy event of IceCube, $E_d=13$~PeV,  and  $E_d=30$~PeV to highlight the  perspective.
We note that the curves  approach asymptotic values of  $\kappa$ as the  distance of the neutrino source, $z_e$, increases towards the  horizon of the Universe.
One  can  see  that merely a single  13~PeV  event constrains the neutrino mass to values much below 1~eV, i.e., near the KATRIN limit, for cosmological sources in the considered (reasonable)  range of $z_e$ values.
In the passing, even so near a  galaxy like NGC~1068 alone ($z_e=0.0038$) delivers quite a tight, independent  bound on the tachyonic neutrino mass of about $5\div 6$~eV, as inferred from  the contour represented by  the  separate dashed line in~Fig.~\ref{fig:Fig6}.

\begin{figure}[!htbp]
    \begin{center}
        \includegraphics[width=0.9\linewidth]{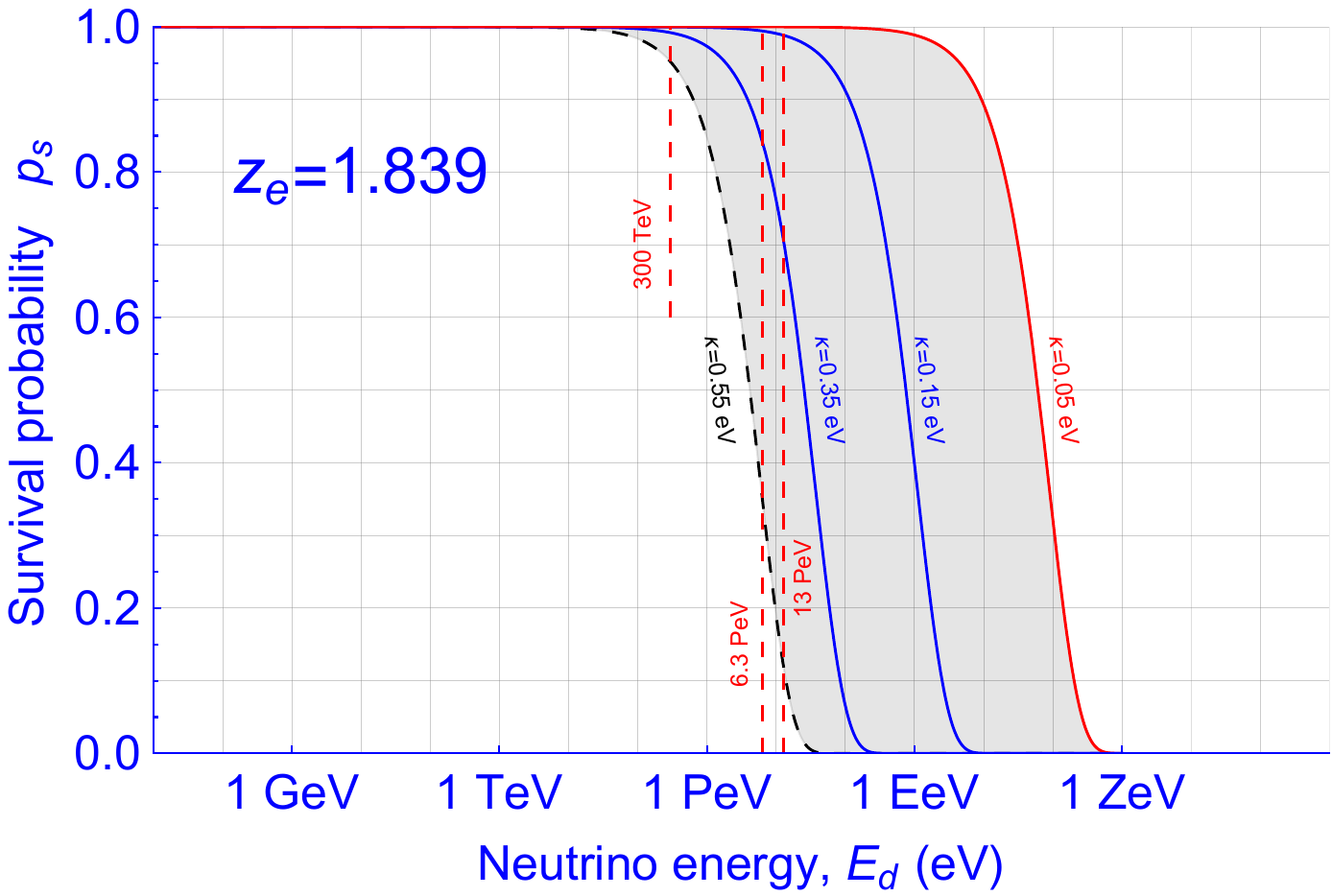}
    \caption{\label{fig:Fig4} Survival probability as a function of tachyonic neutrino energy for $z_e=1.839$ (PKS 1502+106). Black dashed curve shows the prediction  for the KATRIN limit of  $\kappa=0.55$, solid curves are drawn for $\kappa$=0.35, 0.15 and 0.05~eV. Red vertical dashed lines mark energies of 300~TeV,  6.3~PeV and 13~PeV. The shaded area marks the allowed region for the tachyonic neutrino mass, between the KATRIN limit and the minimal value.}
    \end{center}
\end{figure}

\begin{figure}[!htbp]
    \begin{center}
        \includegraphics[width=0.9\linewidth]{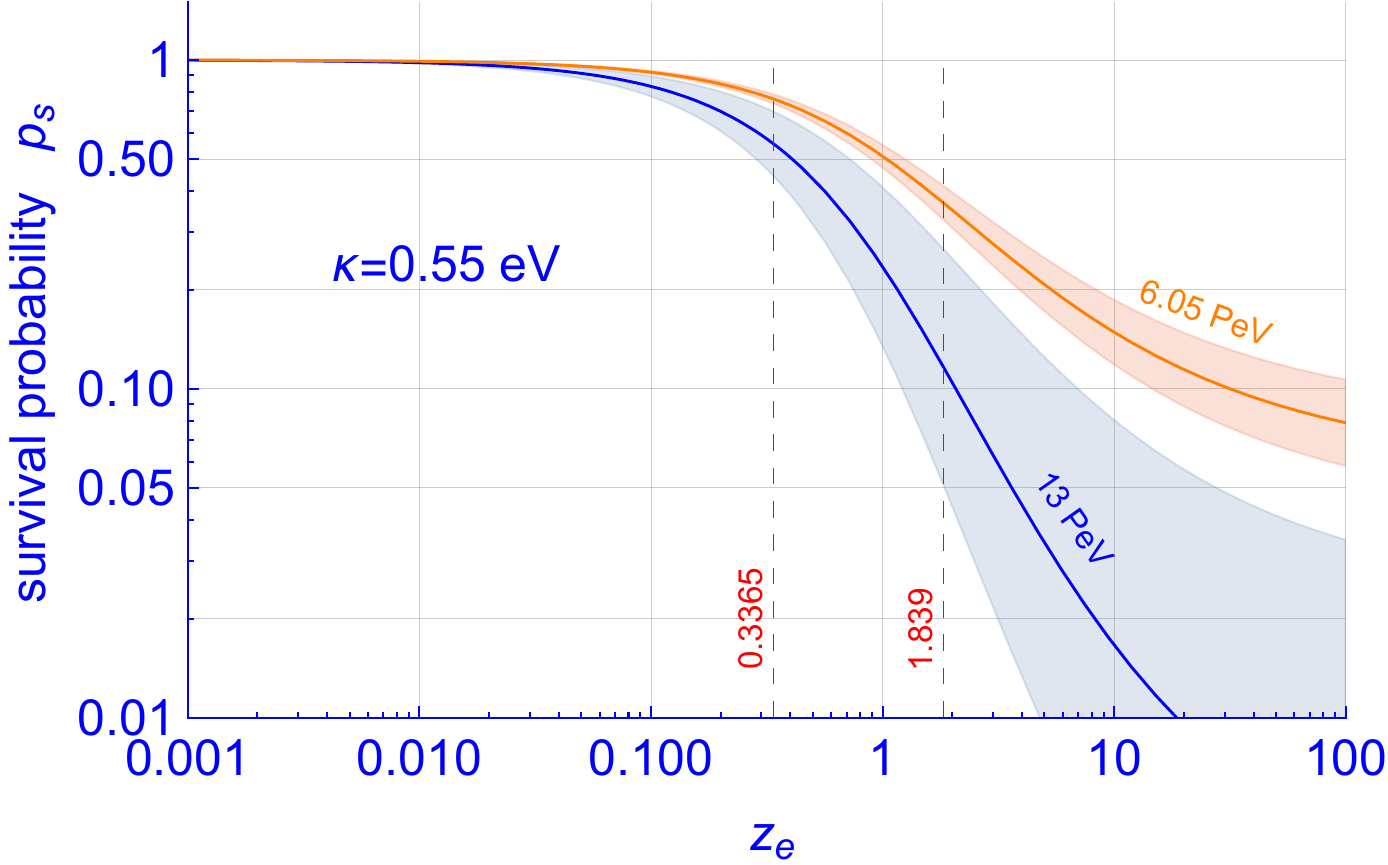}
    \caption{\label{fig:Fig5}  Survival probability  vs. $z_e$ for  $\kappa=0.55$~eV (KATRIN limit) and  energies  $E_d=6.05\pm 0.72$~PeV (Glashow resonance -- measured values) and   $E_d=13\pm 5$~PeV (highest energy event in IceCube). The shaded bands represent the range of the respective   experimental uncertainties.}
    \end{center}
\end{figure}

Survival probability as a function of the observed neutrino energy is shown in Fig.~\ref{fig:Fig4} for a fixed  value of $z_e$=1.839 (PKS 1502+106). The dashed black  curve corresponds to $\kappa=0.55$~eV  (KATRIN limit) and yields $p_s=0.95$ for this 300~TeV event (for the blazar TXS 0506+056  one obtains  $p_s=0.99$). Thus one can  conclude that  the energy spectra  from these two sources (should the statistics increase in the future) would be  practically undistorted by decays  for the range of tachyonic neutrino masses  below  the KATRIN limit.
It can be also seen in Fig.~\ref{fig:Fig4}  that even  at this mass  limit the  survival probability of  the  neutrino (from this assumed distance) with the  energy of the Glashow resonance, $p_s\approx$0.33, is still  high enough to justify observing one  event.
Dependence of the survival probability   on $z_e$   for  two energies, the Glashow resonance energy  and for   13~PeV, adopting  $\kappa=0.55$~eV, is shown in Fig.~\ref{fig:Fig5}. In the  former case, neutrinos originating  even from  distant sources up to $z_e \approx  10$, are still likely to survive the  way to   Earth, while in the latter one would expect such neutrinos to predominantly decay if emitted from distances larger than that of the blazar PKS 1502+106 (for the above value of $\kappa$).

\begin{figure}[!htbp]
    \begin{center}
        \includegraphics[width=0.9\linewidth]{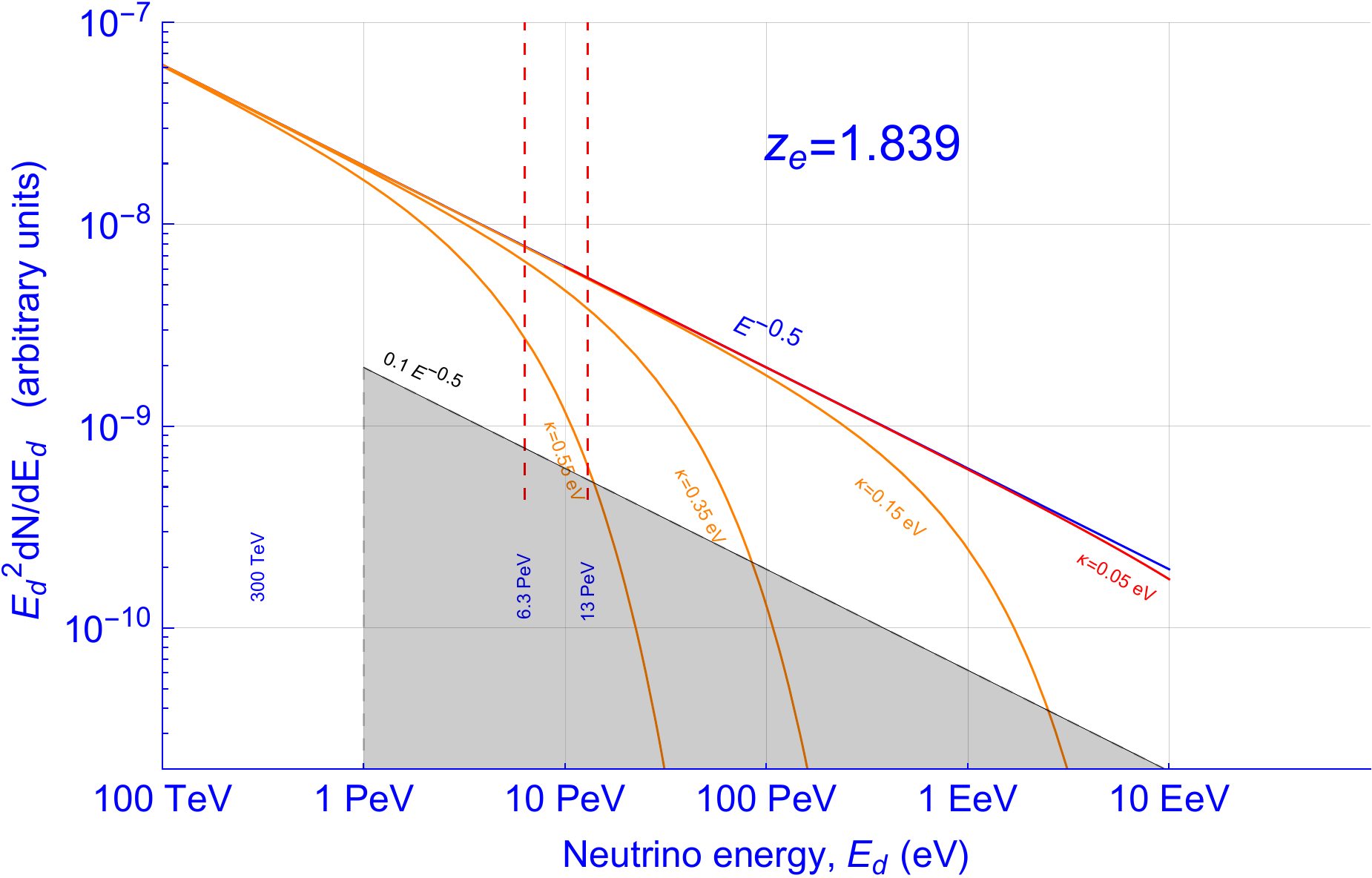}
    \caption{\label{fig:Fig7f} Effect of  decays on the neutrino   energy spectrum, $E_d^2 dN/dE_d \propto E_d^{-0.5}$    from  an assumed  source at  $z_e=1.839$~(PKS 1502+106).
    The curves are drawn for the indicated values of  neutrino mass, including the lowest possible value, $\kappa=0.05$~eV. The shaded area marks the region of fluxes depleted by more than one order  of magnitude. }
    \end{center}
\end{figure}

Now, the most suggestive manifestation of neutrino decays would be a depletion of the energy spectrum of cosmological neutrinos, $dN/dE_d$,  near the high-energy end, as suggested by observations~\cite{cite:Anchordoqui2017}. An  illustration of this effect is shown in Fig.~\ref{fig:Fig7f}. The neutrino energy spectrum, without
referring to any particular type of the  source, is drawn assuming a commonly used parametrisation of the   energy dependence,   $dN/dE_d \propto E_d^{-\gamma}$, where 
$\gamma = 2.5$ (straight line).  The curves represent depleted spectra, obtained as the product of  $dN/dE_d$ with  the survival probability~(\ref{eq:CosmicProb3}), for a range of values of
$\kappa$,  with the redshift of the hypothetical source fixed to $z_e=1.839$~(PKS 1502+106). The shaded area marks the region where the flux is depleted by more than 
an order of magnitude.  One can see that the flux of neutrinos with energies  100~PeV or more   would already be strongly depleted for $\kappa>0.35$~eV; similarly, the 
neutrino flux  at 1~EeV  would suffer from decays for $\kappa$ as low as 0.15~eV (for a nearer source one would expect a similar  effect however of  a smaller 
magnitude). Moreover, should one experimentally  observe a depletion on the basis of a higher statistics data in the future or, similarly,  confirm a non-observation 
 of the ultra-high energy neutrino flux above, one could   attempt to determine  a lower limit of the neutrino mass.
We finally note that if the value of the smallest   neutrino mass state is very  near zero,  implying   the mass of the heaviest state  equal 0.05~eV, the neutrinos can be considered stable over the entire visible Universe for energies even  up to about 10~EeV.

Until now  we have  considered  the decay  only in the aspect of  disappearance of the primary neutrino. However  the decay shown in  Fig.~\ref{fig:Graph1}
is a $1 \rightarrow 3$ process,  in which the energy  of the  initial neutrino is shared, in general unequally, among the three secondaries which   are emitted at non-zero angles w.r.t.  the momentum vector of the primary neutrino. This fact may have  consequences for a detailed  picture of high and ultra-high energy cosmic neutrinos.
Neutrinos which are emitted from the source into the solid angle covering of  the Earth and do  not decay,   travel   along a straight line connecting  the point of emission (source)  and the point of interaction in the detector (Earth). The time-of-flight  of such an ultra-high energy neutrino is practically equal to that of a photon, so if a neutrino and a photon  are emitted simultaneously at the source, both will reach the Earth in coincidence.
As for the   neutrinos  which  potentially  do  decay,  all   three  secondaries arising at a random location on the line of sight   subsequently  miss  the Earth. Consider however    neutrinos which are not emitted into the solid angle of the Earth  which   subsequently  decay. It may happen that one of the   secondaries arising  at some angle w.r.t. its direction of flight,    enters   the  solid angle of the detector on Earth.
Such neutrinos  can still have very high  energies but travel  a longer distance (time) and can drop off from the coincidence with the electromagnetic component.
Assessing whether this is a significant effect requires however a dedicated study of  differential distributions $d \Gamma /d \Phi$~(\ref{eq:dGamma1}).


\section{Summary and conclusions}

We have analysed the decays of high and ultra-high energy neutrinos under an assumption  that these particles  are spacelike  -- a hypothesis already  standing for some time, inspired  by theoretical and experimental results. It has been shown that not observing  ultra-high energy neutrinos on Earth  can  be  explained as due to decays,   $\nu_{\alpha}\rightarrow \nu_{\alpha} \; \nu_{\beta} \bar{\nu}_{\beta}$, if neutrinos were spacelike fermions.
We have calculated the Standard Model width for such decays,  with the tachyonic neutrino mass being the only quantity not precisely known. Subsequently we derived   an exact  expression for survival probability on the way  to  Earth, parametrised   by the emission energy, $E_e$, or energy measured on Earth, $E_d$, and the redshift of the source, $z_e$,  in the environment of  expanding Universe.

We exploited the fact of observing  five   high neutrino events ($i$)--($v$) to discuss and demonstrate a possibility to set approximate upper limits on the tachyonic neutrino mass.
We have shown that recording  the highest energy 13~PeV event allows  to put an  upper mass limit  of about $0.6\div 0.8$~eV,  subject of  reasonable assumptions.
These approximate values fall very near to  the range allowed by the latest measurement of KATRIN, $\kappa <$0.55~eV at 90\% c.l.   The consistency   of our cosmology-based  estimations  with the results from a terrestrial $\beta$-decay  experiment   is remarkable.
We also show  that  a  neutrino mass in the approximate range  indicated in Fig.~\ref{fig:Fig7f} can explain  the non-observation of ultra-high energy tachyonic neutrinos as due to their  decays,  provided  the neutrino mass is not smaller than  about  0.15~eV.  If the fact of "non-observation" was confirmed experimentally on a  sufficient statistics, this could lead to establishing a lower limit on the neutrino mass --  first result of this kind as regards the mass of the neutrino.
Lastly, the smallness of the neutrino masses  seems to provide some  rationale to the hypothesis of tachyonic neutrinos.  If the neutrino mass amounted to only    a few~eV,   neutrino fluxes from  nearby galaxies would  be depleted at already TeV energies.


\noindent
{\bf Funding}

This research did not receive any specific grant from funding agencies in the public, commercial, or not-for-profit sectors.

\noindent
{\bf Declaration of competing interest}

The authors declare that they have no known competing financial interests or personal relationships that could have appeared to influence the work reported in this paper.

\noindent
{\bf Data availability}

No data was used for the research described in the article.

\section{Acknowledgments}
We wish to thank W. Bednarek for fruitful  discussions regarding cosmic rays and ultra-high energy  neutrinos.


\begin{thebibliography}{00}

\bibitem{cite:Aartsen2017a}  IceCube Collaboration, M. G. Aartsen  et al., The IceCube Neutrino Observatory: Instrumentation and online systems,
J. Instrum. \textbf{12}, P03012, (2017).

\bibitem{cite:Aartsen2013a}  IceCube Collaboration, M. G. Aartsen et al., Evidence for High-Energy Extraterrestrial Neutrinos at the IceCube Detector,
Science \textbf{342}, 1242856 (2013).

\bibitem{cite:Ackermann2015a} M. Ackermann et al., The spectrum of isotropic diffuse gamma-ray emission between 100 MeV and 820 GeV, Astrophys. J. \textbf{799}, 86 (2015).


\bibitem{cite:Aartsen2018a} IceCube Collaboration, M.G. Aartsen et al.,  Neutrino emission from the direction of the blazar TXS 0506+056 prior to the
IceCube-170922A alert, Science \textbf{361} (2018) 147.

\bibitem{cite:Kadler2016a} M. Kadler et al., Coincidence of a high-fluence blazar outburst with a PeV-energy neutrino event, Nat. Phys.\textbf{ 12}, 807 (2016).

\bibitem{cite:Lipunov2019a} V. Lipunov et al., IceCube-190730A: MASTER alert observations and analysis,\\ The Astronomer's Telegram, 12971 (2019); www.astronomerstelegram.org/?read=12971.

\bibitem{cite:Aartsen2020a} IceCube Collaboration, M. G. Aartsen et al., Time-Integrated Neutrino Source Searches with 10 Years of IceCube Data,
Phys. Rev. Lett.\textbf{ 124}, 051103 (2020).
\bibitem{cite:Abbasi2022b} IceCube Collaboration, R. Abbasi et al., Evidence for neutrino emission from the nearby active galaxy NGC 1068, Science \textbf{378}, 538 (2022).

\bibitem{cite:Aartsen2013b} IceCube Collaboration, M. G. Aartsen et al., First observation of PeV-energy neutrinos with IceCube,
Phys. Rev. Lett. \textbf{111}, 021103 (2013).

\bibitem{cite:Aartsen2021a} IceCube Collaboration, M. G. Aartsen et al., Detection of a particle shower at the Glashow resonance with IceCube, Nature \textbf{591},
220 (2021); Publisher Correction: Detection of a particle shower at the Glashow resonance with IceCube, Nature \textbf{592}, E11, (2021).

\bibitem{cite:Abbasi2023b} IceCube Collaboration, R. Abbasi et al., Updated directions of IceCube HESE events with the latest ice model using DirectFit, PoS ICRC2023, (2023);[astro-ph.HE] 2307.13878v1.





\bibitem{cite:Aab2020a} PIERRE AUGER Collaboration, A. Aab, et al., Features of the Energy Spectrum of Cosmic rays above $2.5 \times 10^{18}$~eV Using the Pierre Auger Observatory,  Phys. Rev. lett. \textbf{125}, 121106 (2020) [2008.06488].
\bibitem{cite:Abraham2008a}  PIERRE AUGER Collaboration, J. Abraham et al.,  Observation of the suppression of the flux of cosmic rays above $4\times 10^{19}$~eV, Phys. Rev. lett. \textbf{101}, 061101 (2008).
\bibitem{cite:Anchordoqui2019} L. A. Anchordoqui et al., Ultra-high-energy cosmic rays, Phys. Rep. \textbf{801}, 1  (2019).

\bibitem{cite:Greisen1966}  K. Greisen, End to the cosmic ray spectrum?, Phys. Rev. Lett. \textbf{16}, 748 (1966).
\bibitem{cite:Zatsepin1966}  G. Zatsepin and V. Kuzmin, Upper limit on the spectrum of cosmic rays,  JETP Lett. \textbf{4}, 78 (1966).

\bibitem{cite:Berezinsky1969} V. Berezinsky and G. Zatsepin, Cosmic rays at ultra-high energies (neutrino?), Phys. Lett. B \textbf{28}, 423 (1969).
\bibitem{cite:Strecker1979} F. Strecker, Diffuse fluxes of cosmic high-energy neutrinos, Astrophys. J. \textbf{228}, 919 (1979).
\bibitem{cite:Hill1983}  C. T. Hill and  D. M. Schramm, Ultrahigh-Energy Cosmic Ray Neutrinos, Phys. Lett. B \textbf{131}, 247 (1983).
\bibitem{cite:Yoshida1993} S. Yoshida and M. Teshima,  Energy spectrum of ultrahigh-energy cosmic rays with extragalactic origin, Prog. Theor. Phys. \textbf{89}, 833 (1993).

\bibitem{cite:Coleman1997} S. R. Coleman and S. L. Glashow, Cosmic ray and neutrino tests of special relativity, Phys. Lett. B \textbf{405}, 249 (1997) [hep-ph/9703240].
\bibitem{cite:Gorham2012} P. W. Gorham et al., Implications of ultrahigh energy neutrino flux constraints
for Lorentz-invariance violating cosmogenic neutrinos, Phys. Rev. D \textbf{86}, 103006 (2012).
\bibitem{cite:Anchordoqui2014} L. A. Anchordoqui et al., End of the cosmic neutrino energy spectrum, Phys. Lett. B \textbf{739}, 99 (2014).

\bibitem{cite:Kraus2005} C. Kraus et al. (Mainz Collaboration), Final results from phase II of the Mainz neutrino mass search in tritium $\beta$
decay, Eur. Phys. J. C \textbf{40}, 447 (2005).
\bibitem{cite:Aseev2011} V. N. Aseev et al. (Troitsk Collaboration), Upper limit on the electron antineutrino mass from the Troitsk experiment, Phys. Rev. D \textbf{84}, 112003 (2011).

\bibitem{cite:Aker2019} M. Aker, et al., An improved upper limit on the neutrino mass from a direct kinematic method by KATRIN, Phys. Rev. Lett. \textbf{123}, 221802 (2019).
\bibitem{cite:Aker2022} M. Aker et al., Direct neutrino-mass measurement with sub-electronvolt sensitivity, Nature Physics, https://doi.org/10.1038/s41567-021-01463-1 (2022).



\bibitem{cite:Kamoi1971} K. Kamoi and S. Kamefuchi, Comments on Quantum Field Theory of Tachyons, Prog. Theor. Phys. \textbf{45}, 1646 (1971).
\bibitem{cite:Nakanishi1972} N.~Nakanishi, Indefinite-Metric Quantum Field Theory, Progr. Theor. Phys. Suppl. \textbf{51}, 1 (1972).
\bibitem{cite:Chodos1985} A.~Chodos, A.I. Hauser and V.A. Kostelecký, The neutrino as a tachyon, Phys. Lett. B \textbf{150}(6), 431 (1985).

\bibitem{cite:Rembielinski1997} J. Rembieli\'{n}ski, Tachyons and  preferred frames, Int. J. Mod. Phys. A \textbf{12} 1677 (1997).
\bibitem{cite:Rembielinski2021a} J. Rembieli\'{n}ski, P. Caban and J. Ciborowski, Quantum field theory of space-like neutrino, Eur. Phys. J. (2021) \textbf{81}, 716 (2021).


\bibitem{cite:Baumann2019} D.~Baumann et al.,  First constraint on the neutrino-induced phase shift in the spectrum of baryon acoustic oscillations, Nat. Phys. \textbf{15}, 465 (2019).


\bibitem{cite:Chodos1992} A.~Chodos et al., Null experiments for neutrino masses, Mod. Phys. Lett. A \textbf{07}, 467 (1992).
\bibitem{cite:Caban2006} P. Caban, J. Rembieli\'{n}ski and  K. A. Smoli\'{n}ski, Decays of spacelike neutrinos, Concepts of Phys., \textbf{3},  81 (2006); arXiv:hep-ph/9707391.
\bibitem{cite:Ehrlich2015}	R.~Ehrlich, Six observations consistent with the electron neutrino being a $m^2 = - 0.11 \pm 0.02$~eV$^2$ tachyon, Astroparticle Phys. \textbf{66}, 11 (2015).
\bibitem{cite:Somogyi2019}
	G.~Somogyi, I.~N\'andori and  U. D. Jentschura, Neutrino splitting for Lorentz-violating neutrinos: Detailed analysis, Phys. Rev. D \textbf{100}, 035036 (2019).
\bibitem{cite:Schwartz2016} C. Schwartz, Toward a Quantum Theory of Tachyon Fields, Int. J. Mod. Phys. A \textbf{ 31}, 1650041 (2016).



\bibitem{cite:Cohen2011} A. Cohen and S. L. Glashow, Pair Creation Constrains Superluminal Neutrino Propagation, Phys. Rev. Lett. \textbf{107}, 181803 (2011).


\bibitem{cite:NuFit} I. Esteban et al., The fate of hints: updated global analysis of three-favor neutrino oscillations, JHEP \textbf{9}, 178 (2020).

\bibitem{cite:Caban2006a} P.~Caban et al., Oscillations do not distinguish between massive and tachyonic neutrinos, Found. Phys. Lett. \textbf{19}(6), 619 (2006).

\bibitem{cite:KolbTurner} E. Kolb and E. Turner, The Early Universe,  CRC Press (1994).
\bibitem{cite:Narlikar1976} J. V. Narlikar and E. C. G. Sudarshan, Tachyons and Cosmology, Mon. Not. R. Astr. Soc. \textbf{175}, 105 (1976).
\bibitem{cite:Workman2022} R. L. Workman et al., (Particle Data Group), Prog. Theor. Exp. Phys. \textbf{2022}, 083C01 (2022).
\bibitem{cite:Dodelson2020} S. Dodelson, and  F. Schmidt, Modern cosmology, Elsevier, (2020).
\bibitem{cite:Anchordoqui2017} L. A. Anchordoqui et al., Evidence for a break in the spectrum of astrophysical neutrinos, Phys. Rev. D \textbf{95}, 083009  (2017).






\end{thebibliography}
\end{document}